\documentstyle[12pt]{article}
\begin{document}
\begin{titlepage}
\begin{flushright} UTPT-94-07 \\ hep-ph/9403228 \\ March 1994
\end{flushright}
\vspace{24pt}
\begin{center} {\LARGE Anomalous Thresholds} \\ \vspace{10pt}
{\LARGE and the Isgur-Wise Function}
\\
\vspace{40pt} {\large B. Holdom$^{\rm a}$ and M. Sutherland$^{\rm b}$}
\vspace{0.5cm}

{\small Department of Physics, University of Toronto\\ 60 St. George
St., Toronto, Ontario\\Canada M5S 1A7}

\vspace{12pt}

\begin{abstract}
The original de Rafael-Taron bound on the slope of the Isgur-Wise function at
zero recoil is known to be violated in QCD by singularities appearing in an
unphysical region.  To be consistent, quark models must have corresponding
singularity structures. In an existing relativistic quark-loop model, the
meson-quark-antiquark vertex is such that the required singularity is an
anomalous threshold.  We also discuss the implications of another anomalous
threshold, whose location is determined by quark masses alone.

\end{abstract}
\end{center}

\vspace{24ex}
{\small $^{\rm a}$ holdom@utcc.utoronto.ca

$^{\rm b}$ marks@medb.physics.utoronto.ca}
 \end{titlepage}

\newpage

The properties of QCD as the quark masses $m_b , m_c \rightarrow \infty$
\cite{1} imply that the spectrum of the semileptonic decay $\overline{B}
\rightarrow D^* \ell \overline{\nu}$ at the zero-recoil point
$\omega\equiv v_B\cdot v_{D^*} =1$ is absolutely normalized at leading order
and receives no corrections at order $1/m_Q$ \cite{2}.  This
gives rise to the possibility of a precise measurement of $V_{cb}$
\cite{3}, provided {\it a)} the corrections beginning at order $1/m_Q^2$ are
proven to be negligible, and {\it b)} a way can be found to extrapolate from
the data at nonzero recoil $\omega>1$ back to the zero-recoil point, where the
rate vanishes kinematically.  The latter requires knowledge of the shape of
the Isgur-Wise function $\xi(\omega)$ which determines the spectrum at
leading order and satisfies $\xi(1)=1$.  [In practice, it also requires
knowledge of the magnitude and shape of the higher-order corrections, which
are nonvanishing away from zero recoil even at order $1/m_Q$.]

In this letter we wish to look at the leading-order problem,
namely the shape of the Isgur-Wise function.  Some time ago, de Rafael and
Taron
derived a lower bound on its slope at zero recoil \cite{dRT}, the value of
which is $\xi^{\prime}(1)\geq -0.89$ \cite{dRT2}. This bound has since been
shown to
depend on assumptions which are not true in QCD.  The bound also seems to be at
odds
with the data which favors a slope somewhat more negative than $-1$.  We
view any significant violation of the bound as providing an interesting clue to
the underlying physics.

In particular it was pointed out by several groups \cite{pointersout} that the
original derivation assumed that the $b$-number form
factor $F(q^2)$ defined by
\begin{equation} \langle \overline{B}(p^{\prime})|\overline{b}\gamma_{\mu}b|
\overline{B}(p) \rangle = F(q^2) (p+p^{\prime})_{\mu} \label{aa} \end{equation}
is analytic in the region below the $B\overline{B}$ threshold at $q^2=4M_B^2$.
The singularities corresponding to the three $\Upsilon$'s which lie just below
threshold were ignored.  These singularities can cause
$\partial F /\partial q^2$ to become more positive at $q^2=0$. A quantitative
estimate of the contribution of these singularities to the slope requires
knowledge of the couplings of the $\Upsilon$'s to the vacuum and the
$B\overline{B}$ pair, and involves large uncertainties
\cite{dRT2,pointersout}.

In the heavy-quark limit $m_b\rightarrow
\infty$, the Isgur-Wise function is related to $F$ by
\begin{equation} \xi(\omega) = F(2M_B^2(1-\omega)) \end{equation} so the effect
of the singularities is to allow
$\xi^{\prime}(\omega)$ to become more negative at
$\omega=1$.  In the following, we will use $\omega$ as the variable.  The
$B\overline{B}$ threshold occurs at $\omega=-1$, and the original derivation of
the bound assumed that $\xi(\omega)$ is analytic for all $\omega>-1$.

The authors of the first three papers of Ref. \cite{pointersout} also
pointed out that it is possible to construct an example of a meson for which
the slope would be large and negative even without the contributions of the
$\Upsilon$ states.  This is an artificial $B$ meson for which the two quarks
are unconfined and are bound in a Coulomb potential.  In this case the mass of
the $B$ meson is greater than the sum of the masses of the two
quarks.  This produces a singularity due to an anomalous threshold
in the region $-1<\omega<1$, which leads to a violation of the bound.

The present authors have recently developed a relativistic quark model
of heavy-light mesons \cite{simply,corrections}.  When expanded in inverse
powers
of the heavy-quark masses the model is consistent with all constraints imposed
by QCD via the heavy-quark effective theory, a property not shared by other
popular quark models \cite{Neurevue}.  The Isgur-Wise function obtained in the
heavy-quark limit of the model violates the de Rafael-Taron bound, and we wish
to
make clear how this can occur in a model of this type.  This result may be
contrasted with the simple quark loop model briefly described by de Rafael
and Taron \cite{dRT,dRT2}, which happens to satisfy their bound.

The mechanism by which the bound is violated in our model is somewhat
analogous to the artificial $B$ meson mentioned above. The model
contains a quantity which acts like a mass in the relevant three-point loop
graph, and the sum of this mass and the heavy-quark mass is
greater than the $B$-meson mass.  The result is again a violation of the bound
due to an anomalous threshold in the region $-1<\omega<1$.

The model represents the matrix element of Eq. (\ref{aa}) by a three-point
quark
loop graph with standard propagators for the quarks. The
meson-quark-antiquark vertices contain damping factors proportional to
\begin{equation}
\frac{1}{-k^2+\Lambda^2-i\varepsilon},\label{damping} \end{equation} where
$\Lambda<<m_b$ and
$k$ is the momentum carried by the light quark.  These factors act to suppress
the flow of momenta larger than $\Lambda$ into the light degrees of freedom,
and the condition $\Lambda<<m_b$ reflects the basic physical fact that
the typical momenta in the light degrees of freedom are much smaller than the
heavy
quark mass.

The graph for the matrix element of Eq. (\ref{aa}) contains the product
of a vertex factor and a light-quark propagator carrying the same momentum,
which may be separated using partial fractions as
\begin{equation}
\frac{1}{-k^2+\Lambda^2-i\varepsilon}\frac{1}{-k^2+m_q^2-i\varepsilon}=
\frac{1}{\Lambda^2-m_q^2}\left( \frac{1}{-k^2+m_q^2-i\varepsilon} -
\frac{1}{-k^2+\Lambda^2-i\varepsilon} \right). \end{equation}  The graph
thus decomposes into a sum of graphs, in some of which the light-quark
mass $m_q$ is replaced by $\Lambda$.  In the model the meson mass $M_B$ is
determined from the two-point function and it automatically satisfies the
inequalities
$m_Q+m_q<M_B<m_Q+\Lambda$.  Thus $\Lambda$
can play the role described above.  As illustrative values we use $m_q \simeq
250$ MeV and
$\Lambda \simeq 670$ MeV, where the latter is determined from a fit to the
$B$, $B^*$, $D$, and $D^*$ masses \cite{simply,corrections}.

Anomalous thresholds occur in the three-point graphs at values of $\omega$ for
which {\it a)} all three particles internal to the graph are on their mass
shell and {\it b)} the velocity of the heavy quark after interaction with the
current is such that it can re-combine with the light degrees of freedom to
form the final state meson.  We have
\begin{equation} \omega_{\rm anom}
=1+\frac{[M_B^2-(m_b+m)^2][M_B^2-(m_b-m)^2]}{2M_B^2m^2},\label{equal}
\end{equation} where $m$ can be either $m_q$ or $\Lambda$ and is independent
of $m_b$.  In the $m_b\rightarrow \infty$ limit, we may write $M_B=m_b+
\overline{\Lambda}$, where $\overline{\Lambda}$ is the leading-order mass
difference between the meson and the heavy quark.  [It is calculable in the
model in terms of $\Lambda$ and $m_q$, and for the above values
of the latter we find $\overline{\Lambda}\simeq 500$ MeV.]

The anomalous threshold corresponding to $m=\Lambda$ is located at
\begin{equation}
\omega_{\Lambda}= 2\frac{\overline{\Lambda}^2}{\Lambda^2} -1 \label{lamthresh}
\end{equation}  In the model, $\overline{\Lambda}$ is always less
than $\Lambda$, so this anomalous threshold occurs
in the region $-1<\omega<1$.  We could
refer to this as a fake anomalous threshold since there is not a set of real
particles going on shell.

Another anomalous threshold corresponds to $m=m_q$ and is located at
\begin{equation}  \omega_{m_q}=
2\frac{\overline{\Lambda}^2}{m_q^2} -1
.\label{wcritprime}
\end{equation} $\overline{\Lambda}$ is always greater than
$m_q$ in the model, so the threshold occurs for $\omega>1$.  Its location is
model-independent in the sense that it depends only on the heavy-
and light-quark masses (since $\overline{\Lambda}=M_B-m_b$).  It occurs in any
model with unconfined quarks.

The Isgur-Wise function is completely determined in our model by the
dimensionless ratio $m_q/\Lambda$, which can take values between 0 and 1.
We find that the slope at zero recoil satisfies
$\xi^{\prime}(1)<-1.25$ for all values of $m_q/\Lambda$, thus
violating the de Rafael-Taron bound.   We show in Fig. 1 a plot of
$-\xi^{\prime}(1)$ versus
$m_q/\Lambda$.   For the illustrative values of $m_q$ and
$\Lambda$ quoted above we find $\xi^{\prime}(1)=-1.28$, and the slope is
insensitive to variations in $m_q/\Lambda$ over a considerable region around
this point.

Fig. 2 shows $\xi(\omega)$ in the region
$-1\leq \omega \leq 1$ for our illustrative value of $m_q/\Lambda$. The
singularity lying between $\omega=0$ and 1 is the $\Lambda$ anomalous
threshold.  The singularity at $\omega=-1$ corresponds to the
$B\overline{B}$-threshold.  In Fig. 3 we plot $\xi(\omega)$ for $\omega \geq
1$ to show the appearance and location of the $m_q$-threshold.  It is
interesting
to contrast this step-function type singularity to the pole-like singularity
occurring below $\omega=1$.

When $m_q/\Lambda \rightarrow 0$ the $m_q$-threshold moves off to
$\omega=\infty$, but $\overline{\Lambda}/\Lambda \rightarrow 0.72$ and the
$\Lambda$-threshold remains near $\omega=0$.  Since the bound is still violated
in this limit, this suggests that the
$\Lambda$-threshold and not the
$m_q$-threshold plays the dominant role in the violation of the bound.  For
$m_q/\Lambda \rightarrow 1$ we find $\overline{\Lambda}/\Lambda
\rightarrow 1$. The two anomalous thresholds move together towards
$\omega=1$, and the slope of the Isgur-Wise function becomes very large and
negative.

That the anomalous thresholds are intimately related to the slope of the
Isgur-Wise function is also illustrated by an example described in
\cite{simply}.  There $\overline{\Lambda}$ was artificially set to zero holding
everything else fixed, which corresponds to taking the heavy meson off-shell to
the
point $p^2_B=m^2_b$.  In this case the resulting slope was found to satisfy the
de
Rafael-Taron bound.  In fact the whole Isgur-Wise function reduces to a sum of
two
simple functions, one of which also appeared in the simple de Rafael-Taron
quark model.  In this artificial case, both anomalous thresholds coincide with
the
$B\overline{B}$-threshold at $\omega=-1$ [as seen in Eqs. (\ref{lamthresh}) and
(\ref{wcritprime})], and thus no longer lead to a violation of the bound.

The foregoing discussion has been independent of any particular semileptonic
decay, since the Isgur-Wise function $\xi(\omega)$ is universal.  But for a
given decay, there is a finite physical region from $\omega=1$ to
$\omega_{\rm max}=(M_1^2+M_2^2)/2M_1M_2$, where $M_1$ and $M_2$ are the
initial and final meson masses.  Depending on the value of $m_q/\Lambda$, the
$m_q$-threshold could potentially be in this physical region.  For the
preferred values of
$m_q$ and $\Lambda$ quoted above, however,
$\omega_{m_q}\simeq 7$ which is far beyond the physically-accessible
endpoint $\omega_{\rm max}\simeq 1.6$ for
$\overline{B}\rightarrow D \ell \overline{\nu}$.  We find that the
$m_q$-threshold would occur in the physical region for this decay only if
$m_q/\Lambda > 0.8$. The slope of the Isgur-Wise function would then be
less than $-2.3$, in disagreement with the data.   We see that physically
reasonable results are obtained in the model as long as
the anomalous thresholds do not lie too close to the physical region.

This observation is not surprising due to the fact that the model does not
incorporate
confinement (other than simply dropping imaginary parts of amplitudes).  The
model can
be expected to give reasonable results in physical regions only if it is true
that
confinement plays little role in determining the values of such quantities.
This
would clearly not be the case if the anomalous thresholds, which reflect the
presence
of free quarks, showed up in physical regions.

We may consider physical processes away from the heavy quark limit and ask for
what quark masses will the
$m_q$-threshold lie outside the physical region.  This leads to a more general
three-point function with a meson with mass $M_1$ and heavy quark
$m_1$ going to a meson with mass $M_2$ and heavy quark $m_2$, with the third
particle
in the loop having mass $m$.  For a given pair of mesons the location of the
$m_q$-threshold depends only on quark masses (with $m=m_q$) and it
occurs at
\begin{equation} \omega_{\rm anom} =
\frac{1}{4M_1M_2m^2} \left\{ (M_1^2-m_1^2+m^2)(M_2^2-m_2^2+m^2) +
\Delta^{1/2}  \right\} \label{wanom} \end{equation} where
\begin{equation}
\Delta=
[M_1^2-(m_1+m)^2][M_1^2-(m_1-m)^2][M_2^2-(m_2+m)^2][M_2^2-(m_2-m)^2].
\end{equation}  [Eq. (\ref{equal}) is a special case of Eq. (\ref{wanom}).]
{}From
this we find for the decay
$\overline{B}
\rightarrow D \ell \overline{\nu}$ the
allowed region in the $m_b$-$m_c$-plane shown in Fig. 4, for $m_q=250$ MeV. The
allowed region includes most of the presently favored range for the
quark masses.  The allowed region is even larger for the decay
$\overline{B}\rightarrow D^{*}\ell\overline{\nu}$.

In other decays it is possible that the quark masses are such that an
anomalous threshold does lie in a physical region, in which case our model
would
offer a poor description.  Confinement would at least smooth out the structure
illustrated in Fig. 3, and possibly obliterate it without leaving a trace.  The
latter phenomenon apparently occurs in the solvable two-dimensional 't Hooft
model.  In this model the meson masses may be calculated exactly in terms of
the
quark masses, thus determining the location of would-be anomalous thresholds.
But in Fig. 6 of \cite{swan} no observable effect of such
thresholds appears in the Isgur-Wise function.

In the real world we may consider decays having larger $\omega_{\rm max}$, such
as $B\rightarrow\rho\ell \nu$ for which $\omega_{\rm max}\simeq 3.5$.  The
existence of an anomalous threshold in the physical region depends strongly on
the effective light-quark mass.  Inserting
$m_1=4800$ MeV  and $m=m_2=250$ MeV in Eq. (\ref{wanom}) yields
$\omega_{\rm anom}=4.7$, while $m=m_2=330$ MeV gives $\omega_{\rm anom}=2.3$.
The latter is well inside the physical region.  The existence of some remnant
of
such a threshold appearing in the data remains an intriguing, although slight,
possibility.

In summary, the violation of the de Rafael-Taron bound constrains the
singularity
structure of models of heavy meson decay.  We have shown how
the required singularities can be produced in a model of unconfined quarks.
The particular model we have studied gives reasonable results (i.e. confinement
need not play an important role) as long as the singularities are well outside
the physical regions.   The singularity structure is related to the form of the
meson-quark-antiquark vertex factor in Eq. (\ref{damping}).  We therefore
conclude that in models of this type the vertex factor is
constrained by the requirement that the Isgur-Wise function be steep enough at
zero recoil.  Exponential vertex factors, for example, would fail to provide an
anomalous threshold analogous to the
$\Lambda$-threshold, and the Isgur-Wise function would be correspondingly
shallow at $\omega=1$.

\newpage
\noindent {\bf Acknowledgement}
\vspace{1ex}

This research was supported in part by the Natural Sciences and Engineering
Research Council of Canada.

\newpage
\noindent{\bf FIGURE CAPTIONS} \vspace{6ex}

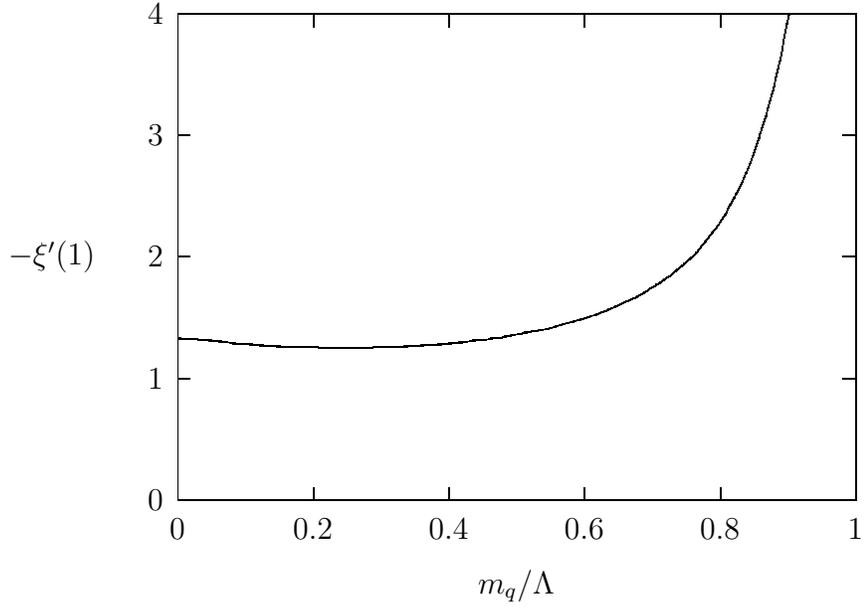
\noindent {\bf FIG. 1:} Negative slope of the Isgur-Wise function at zero
recoil
as a function of the ratio of the effective mass
$m_q$ of the light degrees of freedom to the scale $\Lambda$ of typical loop
momenta.
\vspace{4ex}

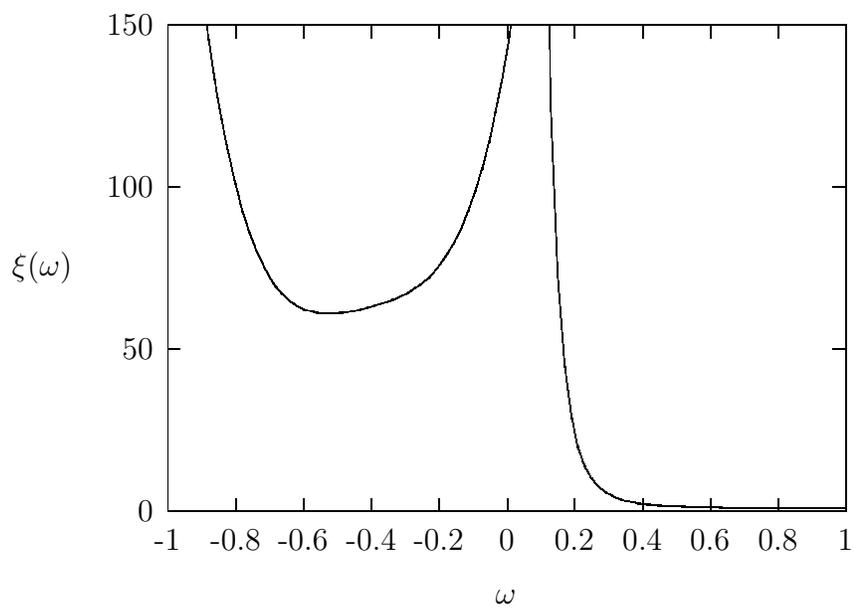
\noindent {\bf FIG. 2:}  Pole-like anomalous threshold in the region $-1 <
\omega < 1$ which makes the slope of the Isgur-Wise function at $\omega =1$
more
negative than the de Rafael-Taron bound.
\vspace{4ex}

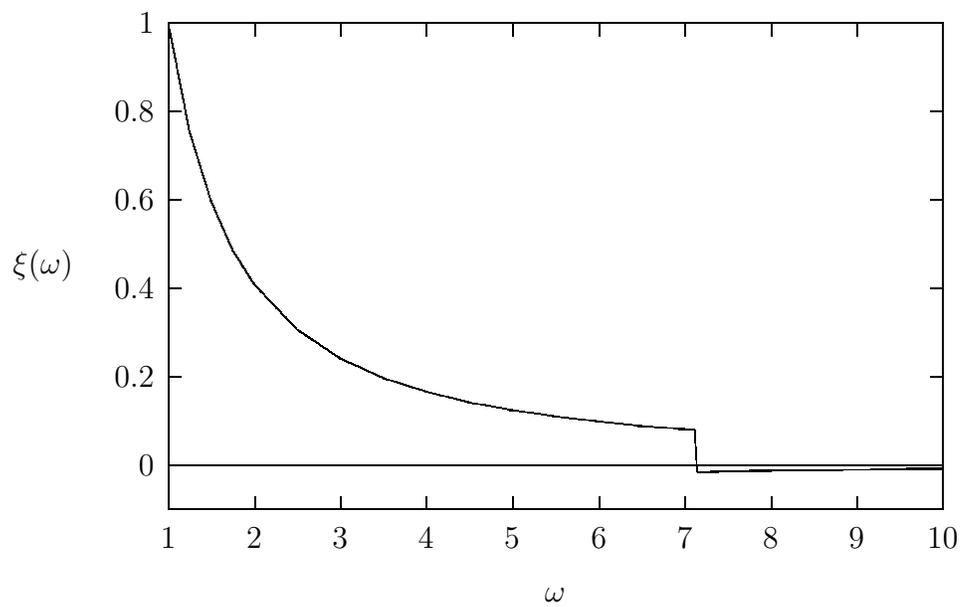
\noindent {\bf FIG. 3:}  Step-like anomalous threshold in the Isgur-Wise
function above $\omega=1$.  [Note: the vertical scale is different from that of
Fig. 2.]
\vspace{4ex}

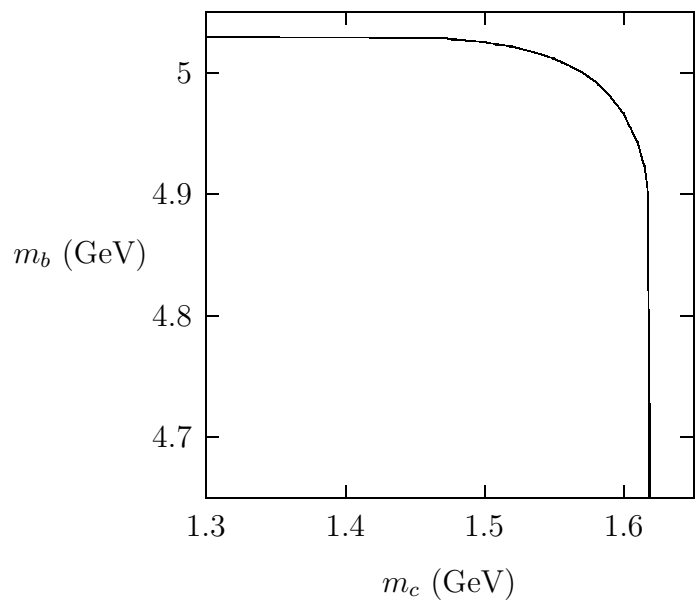
\noindent {\bf FIG. 4:}  Region in ($m_c,m_b$)-space [below and to the left of
the curve] for which the upper anomalous threshold lies above the maximum
recoil
point for $\overline{B}\rightarrow D\ell\overline{\nu}$.

\end{document}